\documentclass[12pt]{article}
\usepackage{epsfig}
\begin{document}

\large
\hfill\vbox{\hbox{DCPT/05/132}
              \hbox{IPPP/05/66}}
\nopagebreak
\vspace{2cm}
\begin{center}
\LARGE{\bf Exclusively Exclusive Final States \\
 in Two Photon Collisions 
}
\vspace{8mm}

\large{M.R. Pennington}
\vspace{4mm}

{Institute for Particle Physics Phenomenology,\\ Durham University, Durham DH1 3LE, UK}

\end{center}
\vspace{5mm}
\small

\begin{abstract}
The study of exclusive final states in two photon collisions
is motivated by the range of physics that can be explored
from chiral dynamics, to resonance physics to quark dynamics,
all within a few GeV of threshold.

\end{abstract}

\normalsize
\baselineskip=5.5mm
\parskip=2.mm

\section{Why Exclusive Channels}
 Two photon interactions are most commonly studied using  $e^+e^-\to e^+e^- X$
without tagging~\cite{brodsky}. The majority of  electrons and positrons are scattered through very small angles, and so the virtual photons that are exchanged  are almost on-shell. The reaction then involves the collision of two essentially real photons. For exclusive channels the final state  $X$ is
typically two to six mesons or a baryon-antibaryon pair.
\begin{figure}[t]
\begin{center}
~\epsfig{file=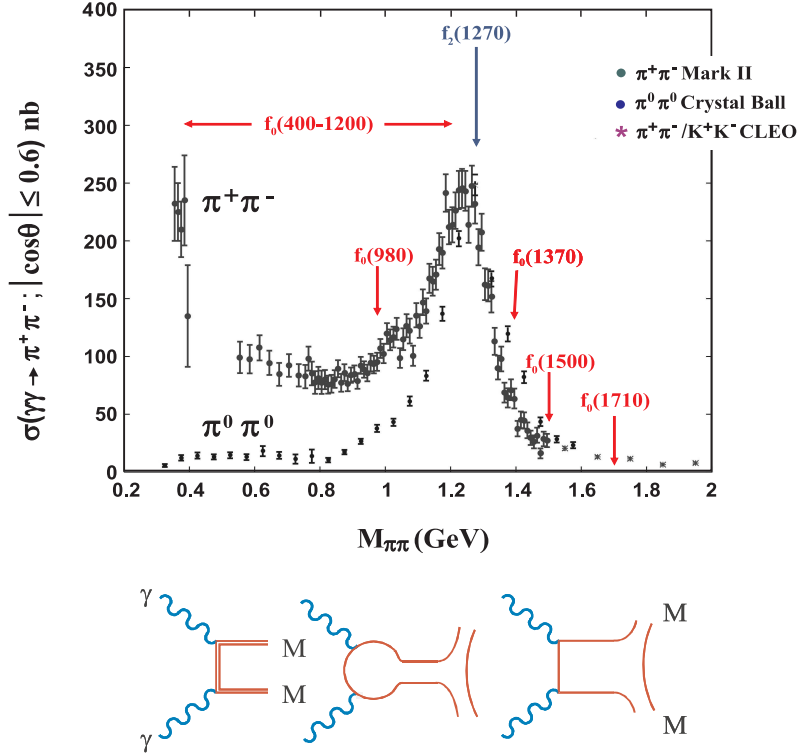,height=10.cm,width=10.5cm}
\caption
{Integrated cross-section for $\gamma\gamma\to\pi\pi$ as a function of c.m. energy $W$ from Mark II~\protect\cite{MarkII}, Crystal Ball (CB)~\protect\cite{CB} and CLEO~\protect\cite{cleo}. The $\pi^0\pi^0$ cross-section has been scaled to the same angular range as the charged data and by an isospin factor for the $f_2(1270)$ peak. Below are graphs describing the dominant dynamics in each kinematic region,
as discussed in the text. } 
\end{center}
\vspace{-7mm}
\end{figure}

To see what these channels teach us, let us consider the cross-section for $\gamma\gamma\to MM$ as a function of the $\gamma\gamma$ c.m. energy $W$. A typical example, where the meson $M$ is a pion either charged or neutral, is shown in Fig.~1. There one sees the cross-section rise from threshold, then have structure and subsequently decline. This cross-section naturally divides into three kinematic regions which correspond to three different dynamical regimes.
In each case the photon couples to the electric charge of a point-like object, but what it sees as point-like  changes with energy. At low momentum, close to threshold, the photon has long wavelength and sees the whole of the final state hadron and couples to its electric charge. It sees the charged pion but not the neutral. This region teaches us about
{\it hadron dynamics}. As the energy increases and the wavelength of the photon shortens, it sees the charged components of the hadron, the constituent quarks. Coupling to them, it causes them to resonate and we learn about {\it resonance dynamics}. The arrows in Fig.~1 point to the resonances listed in the PDG Tables~\cite{PDG}. Lastly, as the energy rises still further, the photon sees charged point-like objects inside the constituent quarks, the current quarks, and we can learn about {\it quark dynamics}. The extent  of the three kinematic regions depends on the final state.
For pions, the three regions are well-separated, for kaons the near threshold region is foreshortened and the resonance regime more structured. In all three cases, the dynamics is governed by QCD. Region~3 is the perturbative--non-perturbative interface, while the lower two depend wholly on the strong physics aspects of QCD. For charmonium and charmed particle production, regions~2 and 3 merge with each other, both being amenable to perturbative treatment.

  Let us discuss region~3, the higher energy regime first. Above a few GeV, when two photons collide in the centre-of-mass frame, they deposit all of their energy in a region of radius a fraction of a fermi. This creates a $q{\overline q}$ pair, which radiate soft gluons, until they reach separations of the order of a fermi, when the
quarks and gluons fragment. This produces two back-to-back jets of hadrons that dominate the inclusive high energy cross-section. To produce an exclusive final state like two mesons, the initial $q{\overline q}$ pair must radiate at least one hard gluon, which in turn creates another $q{\overline q}$ pair, moving parallel to the initial quarks, so that these can get together to form a meson.  It was Brodsky and Lepage~\cite{brodsky-lepage}, who recognised that such processes can be divided into a short distance part governing the emission of the hard gluon and a long distance component determined by the wavefunction of the hadron in the final state. Thus they predicted that the differential cross-section for
$\gamma\gamma\to\pi\pi$ should be given by
\begin{equation}
\frac{d \sigma}{d \cos \theta}\ (\gamma\gamma\to\pi^+\pi^-)\; \simeq \;\frac{8\pi\alpha^2}{W^2}\;\frac{F^2 (W^2)}{\sin ^4 \theta}\; ,
\end{equation}
where $W$ is the energy and $\theta$ the scattering angle in the $\gamma\gamma$ centre-of-mass frame. The meson's annihilation formfactor, $F$, is related to its electromagnetic formfactor~\cite{Diehl}. This prediction, which has long been known to agree with data from Mark II~\cite{MarkII,MarkIIbl}, TPC/Two-Gamma~\cite{TPC} and  CLEO~\cite{cleo},
is shown in Fig.~2 compared with the more recent results from ALEPH~\cite{Aleph} and Belle~\cite{BelleMM} on the production of charged pion and kaon pairs. Eq.~(1) describes the data very well.
\begin{figure}[h]
\begin{center}
~\epsfig{file=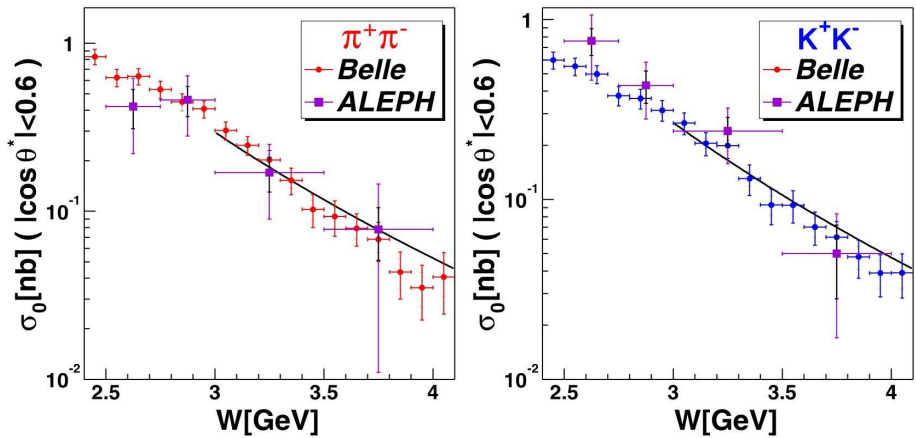,width=12.5cm}
\vspace{-2mm}
\caption
{Integrated cross-section for $\gamma\gamma\to \pi^+\pi^-$ and $K^+K^-$ as a function of c.m. energy $W$ from ALEPH~\protect\cite{Aleph} and Belle~\protect\cite{BelleMM} compared to the predictions of Eq.~(1).} 
\end{center}
\vspace{-2.mm} 
\end{figure}

To form baryons in the final state, a further hard gluon has to be radiated, to produce yet another back-to-back $q$ and ${\overline q}$. Farrar, Maina and Neri~\cite{Farrar} first computed ${\overline p}p$ production by incorporating the  three quark
wavefunction of Chernyak and Zhitnitsky~\cite{Chernyak} for the proton. This makes a prediction  well below the data (see Fig.~3). Though baryons are built of three valence quarks, they do spend part of their time in a quark-diquark configuration. The idea that a scalar diquark, $[ud]$, may be a significant component of the nucleon~\cite{jaffe2} has received renewed interest sparked by the possible discovery of pentaquark baryons. 
With diquarks as intrinsic entities it  becomes natural that baryons
can be produced by hard gluons  creating just one diquark-antidiquark $[ud][{\overline{ud}}]$ system rather than 
two $q{\overline q}$ pairs.  This gives predictions~\cite{diquarkmodels}, for both the energy and scattering angle dependence, that are in far better agreement with the older data from~\cite{oldBB}.
Data with greater precision published in the past year by Belle~\cite{BelleBB}, shown in Fig.~4, intriguingly hint that agreement with updated diquark predictions~\cite{diquarksBelle} may be a transient phenomenon.
The data may be approaching the three valence quark prediction 
at higher energies. Only data with $W > 4$ GeV can confirm this. 
 
\begin{figure}[t]
\begin{center}
~\epsfig{file=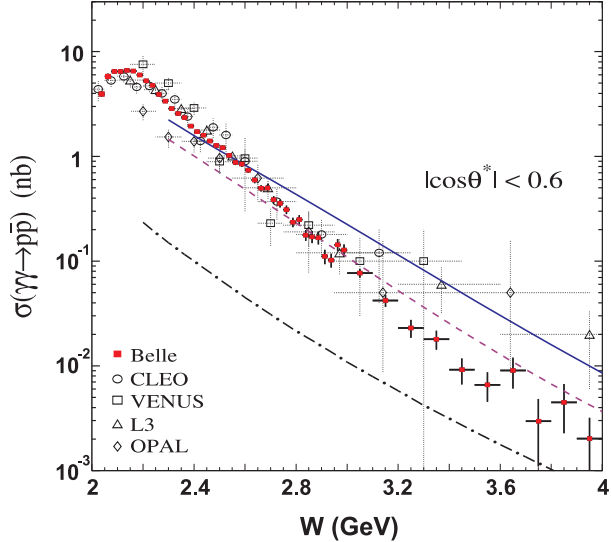,width=8.cm}
\caption
{Integrated cross-section for $\gamma\gamma\to {\overline p}p$ as a function of c.m. energy $W$ from Belle~\protect\cite{BelleBB}, together with the older results from CLEO and VENUS~\protect\cite{oldBB} and the LEP results~\cite{LEPBB} from L3 and OPAL. These are compared with the 3 quark  prediction of Farrar {\it et al.}~\protect\cite{Farrar} (the lowest dashed line) and of the updated diquark models of Berger and Schweiger~\protect\cite{diquarksBelle} (solid  and middle dashed lines).} 
\end{center}
\vspace{-6mm} 
\end{figure}

\begin{figure}[h]
\begin{center}
\includegraphics[width=12.cm,angle=0]{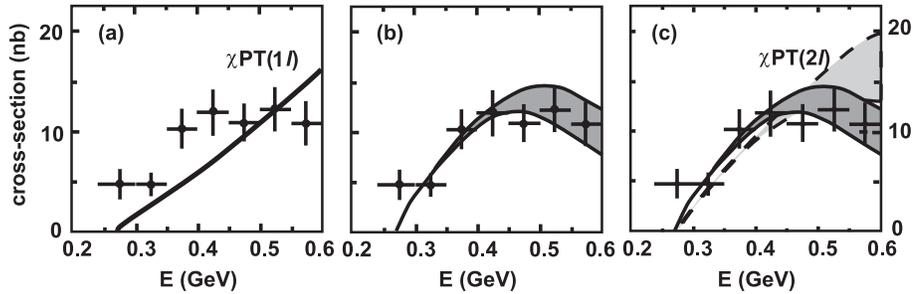}
\vspace{1mm}
\caption{\leftskip 1cm\rightskip 1cm {\small{Integrated cross-section for $\gamma\gamma\to\pi^0\pi^0$ as a function of the $\pi\pi$ invariant mass $E$. The data are from Crystal Ball~\protect\cite{CB} scaled to the full angular range. (a)~the line is the prediction of $\chi PT$ at one loop $(1\ell)$~\protect\cite{bijnens}. (b) shows the dispersive prediction~\protect\cite{mp,penn} --- the shaded band reflects the uncertainties in 
experimental knowledge of both $\pi\pi$ scattering and vector exchanges.
(c) shows the dispersive band of (b) together with the prediction of $\chi PT$ at two loops $(2\ell)$~\protect\cite{bellucci} }}}
\vspace{-6mm}
\end{center}
\end{figure} 

At the lowest energies the photon couples to the whole hadron. There pion interactions are governed by chiral dynamics  embodied in Chiral Perturbation Theory ($\chi$PT). 
The process $\gamma\gamma\to\pi^0\pi^0$ has no Born term, so the lowest order contributions have one loop  and the sum of these graphs is finite. This gives a prediction, shown in Fig.~4(a), which rises almost linearly with energy~\cite{bijnens}, that agrees with the only available data from Crystal Ball~\cite{CB} at just a couple of energies.
Since this process was advertised as a {\it gold-plated test of $\chi$PT}~\cite{maianichpt}, the conclusion by some in the early '90s was that the data must be wrong. However, one can calculate the process non-perturbatively: $\,\gamma\gamma\,$ can go to $\,\pi^+\pi^-\,$  dominated by its one pion exchange Born term 
 at low energies (seen as the near threshold peak in Fig.~1), and then the $\,\pi^+\pi^-\,$ can scatter and go to $\,\pi^0\pi^0\,$ through final state interactions calculable using dispersion relations. These were computed by David Morgan and I~\cite{mp,penn} and are in agreement with experiment, Fig.~4(b). Since chiral dynamics should not be wrong, the problem must be with the  perturbative approximation.
\noindent One loop $\chi$PT includes just tree level $\pi\pi$ interactions and one can easily check that this does not reproduce the experimental  $I=0,2$ $S$-wave phase-shifts~\cite{penn}
that are included in the dispersive calculation. By going to two loop $\chi$PT,
Bellucci, Gasser and Sainio~\cite{bellucci} found agreement with the Crystal Ball data, Fig.~4(c), within the uncertainties (indicated by the shaded band) in the higher order constants. This diffused the need to remeasure this cross-section, at least below 500 MeV.
However, above the threshold region neither the dispersive approach nor higher order $\chi$PT can be reliably computed and we need data. This is not surprising, since only experiment can determine the two photon coupling of resonances, which are such a guide to their composition.

\section{Resonance Dynamics}

  If we now look at the  intermediate energy region with  $\,0.5  < W < 2$ GeV, we see  distinct resonance structures. Tensor mesons always appear strongly in vector-vector interactions. 
In the $K{\overline K}$ channel, the $f_2'(1525)$ and $a_2(1320)$ overlap and interfere, while in the $\pi\pi$ channel the $f_2(1270)$ is seen essentially in tact in Fig.~1. 
Not surprisingly, the relative two photon couplings of these spin-two resonances, $f_2$, $a_2$ and $f_2'$, reflect the fact that they belong to an ideally mixed quark multiplet.
The absolute scale of their couplings depends on dynamics: on how the ${\overline q}q$ pair form the hadron. This is even more the case for the lightest pseudoscalars, $\pi(140)$, $\eta(550)$ and $\eta'(950)$, for which not just the absolute rates but the relative ones depend on the strong dynamics of Fig.~5. For  heavy flavour states, like charmonia, this is reflected simply in the wavefunction at the origin.
\vspace{2mm}
\begin{figure}[h]
\begin{center}
~\epsfig{file=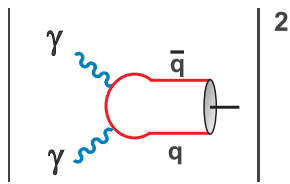,width=3.75cm}
\vspace{0mm}
\caption
{Two photon decay rate of a hadron is the modulus squared of the amplitude for $\,\gamma\gamma\,$ to produce a $\,{\overline q}q\,$ pair and for these to bind by strong coupling dynamics. } 
\end{center}
\vspace{-5mm} 
\end{figure}

Let us now turn to the most enigmatic hadrons of the past forty years: the scalars. They have a special place in strong interaction dynamics, since they 
directly reflect the nature of the QCD vacuum. 
How they couple to $\gamma\gamma$  can help
 to unravel which scalar plays what role in the breaking of chiral symmetry, which is a glueball, which is largely $qq{\overline{qq}}$ and which $q{\overline q}$.  The PDG Tables~\cite{PDG} list with $I=0$ the
 $f_0(400-1200)$ (or~$\sigma$), $f_0(980)$, $f_0(1370)$, $f_0(1500)$ and $f_0(1710)$ below 1.8 GeV, as indicated in Fig.~1.

If all these really exist in the spectrum of scalars (and this is questioned in Refs.~\cite{mrpbeijing,ochs}), then there are enough states to fit into two nonets and still leave one over to be the anticipated glueball. The lightest scalars are very short-lived and so must have large multi-meson components, like $\,\pi\pi\,$ and $\,{\overline K}K\,$, in their Fock space~\cite{vanbeveren}. Perhaps then they are closer in composition to tetraquark states. This idea was proposed long ago by Jaffe~\cite{jaffe} and taken on by Schechter and collaborators~\cite{schechter} amongst others. The resurgence of interest in diquark components has recently made this a popular picture with work by Maiani {\it et al.}~\cite{maiani}. In this model, which should be stressed depends on all the states listed in the PDG Tables being \lq\lq real'', the conventional ${\overline q}q$ nonet, displayed in Fig.~6, is centred around 1400 MeV in mass, with the $a_0/f_0(980)$ the heaviest of the light tetraquark mesons. That the isotriplet is degenerate with the isosinglet that couples strongly to ${\overline K}K$ (as the $f_0(980)$ does) is natural in the 4-quark picture, where these states are orthogonal combinations of $\,{\overline {[sn]}}[sn]\,$  scalar diquarks (where $n=u,d$).
The predicted glueball~\cite{bali}\ would mix primarily with heavier ${\overline q} q$ isoscalars to form components of the \lq\lq observed'' $f_0(1370),\,f_0(1500)$ and $\,f_0(1710)$.

\begin{figure}[h]
\vspace{2mm}
\begin{center}
~\epsfig{file=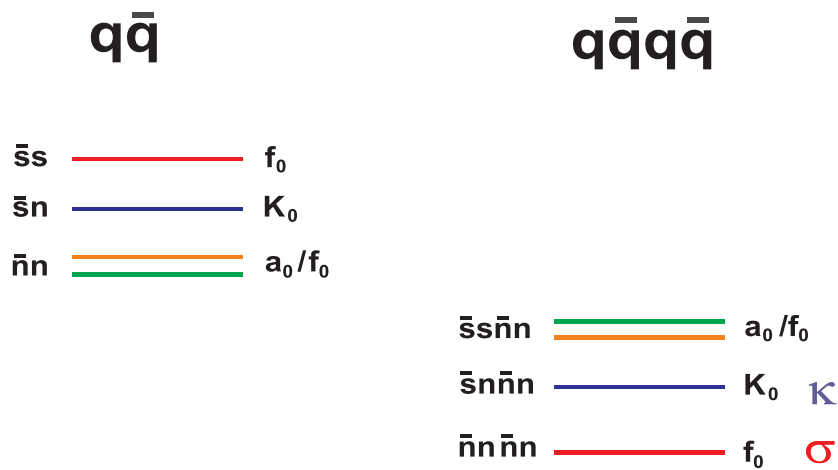,height=6.1cm,width=11.cm}
\vspace{2mm}
\caption{Spectrum of ${\overline q}q$ and ${\overline{qq}}qq$ scalars
in the models of Refs.~\protect\cite{jaffe}-\protect\cite{maiani}.}
\vspace{-4mm} 
\end{center}
\end{figure}

Two photon couplings of these states are the key inputs we need to confirm and clarify this picture. 
This remains a challenge. As shown in Fig.~1 with the $\,\pi\pi\,$ final state, 
the only clear resonance is the $f_2(1270)$. The scalars that  lie underneath this  are more difficult to find.
Only by combining data from both $\pi^+\pi^-$ and $\pi^0\pi^0$ production have we any chance of separating the cross-section into components of definite spin, which is so essential to determining the two photon couplings of the contributing states.
Since we have no polarization information and coverage is at best 80\% of the angular range in $\cos\theta$, we need additional constraints to determine the partial waves. Only for the $\pi\pi$ final state is this possible.  As already remarked $\gamma\gamma\to\pi^+\pi^-$ is dominated by the Born term at low energies, modified by calculable final state interactions. At all energies
unitarity relates the $\gamma\gamma$ process to other hadronic reactions,
on which we may have detailed experimental information.
Below 1.4 GeV or so, when multipion channels are unimportant, just the $\pi\pi$ and $K{\overline K}$ intermediate states are all that need be included. By implementing such constraints, one can make up for the inadequacies of the two photon information by incorporating hadronic scattering data into the codes. 
Following earlier work with David Morgan~\cite{mp2}, Elena Boglione and I~\cite{BP} completed such an Amplitude Analysis of all presently available data. This revealed two classes of solutions with differing  radiative widths for the $f_2(1270)$, $f_0(980)$ and $f_0(400-1200)$, as listed in Ref.~\cite{BP} and the PDG Tables~\cite{PDG}.
Precision two photon cross-sections, differential as well as integrated, can distinguish between these solutions.
CLEO took such data some time ago, but unfortunately these have never been 
finalised. New results from Belle are now eagerly awaited.

Despite this imminent publication, the challenge still remains for one of the $e^+e^-$ colliders around the world  to have a dedicated two photon team committed to delivering accurate measurements, from which the two photon widths of all the low mass scalars can be deduced. This requires the study of all accessible final states, $\pi^0\pi^0$,
$\pi^+\pi^-$, $K^+K^-$, $K^0{\overline K^0}$, $4\pi$, as well as $\pi^0\eta$ to understand the related $I=1$ sector, with as large an angular coverage as possible~\cite{sardinia}. Combining such results with reliable predictions from strong coupling QCD for two photon couplings, we can determine the constitution of these key hadrons. Only then will we understand the nature of the light scalar mesons, a nature and composition that is intimately tied to the structure of the QCD vacuum. They are the Higgs sector of the strong interaction. They may 
serve as a guide to the world of electroweak symmetry breaking awaiting discovery. 
 There two photon processes may also be crucial in providing illumination and exclusive insights.

\vspace{6mm}

\noindent{\bf Acknowledgements}

It is a pleasure to thank Maria Krawczyk and her colleagues for organising this \lq\lq Einstein Year''
 conference highlighting the progress achieved in understanding photons over the past hundred years. We still have much to learn. Partial support of the EU-RTN Programme, 
Contract No. HPRN-CT-2002-00311, \lq\lq EURIDICE'' is acknowledged.

\end{document}